\documentstyle[multicol,aps,prb,epsf]{revtex}
\begin{document}

\draft

\title{
Numerical study of a superconductor-insulator transition 
in a half-filled Hubbard chain with distant transfers
}

\author{
Kazuhiko Kuroki, Ryotaro Arita, and Hideo Aoki
}
\address{Department of Physics, University of Tokyo, Hongo,
Tokyo 113, Japan}

\date{\today}

\maketitle

\begin{abstract}
The ground state of a one-dimensional Hubbard model 
having the next-nearest neighbor hopping $(t')$ as well as the 
nearest-neighbor one $(t)$ is numerically
investigated at half-filling.  
A quantum Monte Carlo result shows 
a slowly decaying 
pairing correlation for a sizeable interaction strength $(U \leq 2t)$, 
while the system is shown to become insulating for yet 
larger $U>U_C\sim 3t$ from a 
direct evaluation of the charge gap with the density-matrix renormalization 
group method. 
The results are consistent with Fabrizio's recent weak-coupling theory 
which suggests a transition from a superconductor into an insulator at a 
finite $U$. 
\end{abstract}

\medskip


\begin{multicols}{2}
\narrowtext

For the past several years, the existence of a gap in the spin excitation
has been suspected to be a key in the 
superconductivity in one-dimensional (1D) strongly correlated systems. 
This view has motivated intensive studies on ladder systems,
\cite{DagRice,Fabrizio,Schulz,Balents,Noack,Kuroki,Kimura} 
and systems with dimerized\cite{Imada} or frustrated\cite{Ogata} spin-spin 
interactions. 
Some of these studies have revealed that
the existence of a spin gap can indeed lead to a dominance of
pairing correlations even in purely repulsive models.
More recently, Fabrizio studied a 1D Hubbard model having 
the next-nearest neighbor (NNN) hoppings in addition to nearest-neighbor (NN)
ones. 
Such a model may have some relevance to the 1D cuprates with large 
ratios of NNN to NN transfers, such as SrCuO$_2$
having a zigzag structure,\cite{Teske} or the recently 
discovered superconductor, 
Sr$_{0.4}$Ca$_{13.6}$Cu$_{24}$O$_{41.84}$\cite{Uehara}, 
where layers of 2-leg ladders alternate with 
layers of 1D chains 
with 90$^{\circ}$ Cu-O-Cu bond angles. 
Making use of a weak-coupling renormalization, 
Fabrizio has predicted the existence of a spin gap 
in a certain parameter regime
and also a `superconductor-insulator
transition' (or more precisely a transition from a metal with dominating
pairing correlation to an insulator since we are talking about 1D systems) 
at a certain Hubbard interaction $U=U_C$ 
at half-filling.\cite{Fabrizio2}  
A transition from a superconductor directly 
into an insulator is intriguing,
but it is beyond the scope of the 
weak-coupling theory to establish 
where (i.e., for which finite value of the interaction strength) 
the transition actually takes place.

The purpose of the present study is to look into the transition 
numerically, and to actually evaluate $U_C$.  
Our strategy here is to employ two complementary approaches: 
the quantum Monte Carlo (QMC) method for small $U$'s and 
the density-matrix renormalization group (DMRG) for large $U$'s.  
Combination of the two has indeed enabled us to establish the 
existence of such a transition at $U_C\sim 3t$.  

The Hamiltonian under consideration is given, in standard notations, as
\begin{eqnarray}
{\cal H}&=&-t\sum_{i \sigma}
(c_{i\sigma}^\dagger c_{i+1\sigma}+{\rm h.c.})
-t'\sum_{i \sigma}
(c_{i\sigma}^\dagger c_{i+2\sigma}+{\rm h.c.})\nonumber\\
&&+U\sum_i n_{i\uparrow}n_{i\downarrow},
\end{eqnarray}
where $t$ and $t'$ are NN and NNN hoppings, respectively, and $U$ is the 
Hubbard repulsion. 
We focus our attention to half-filling, $n=1$, where the signs of $t$
and $t'$ are irrelevant due to an electron-hole symmetry.
Here we shall take $t>0$ and $t'<0$.\cite{Daul}   

Let us first recapitulate Fabrizio's weak-coupling 
theory.\cite{Fabrizio2}
If $|t'/t|$ is small enough, the system can be mapped to 
the $t'=0$ case, 
because there is no essential difference in the vicinity of the 
Fermi level. 
For $|t'|\geq t_c(n)$, on the other hand, 
the Fermi level intersects the one-electron 
band at four $k$-points 
($\pm k_F^1$ and $\pm k_F^2$), 
where $t_c(n)$ is a function of the band 
filling (with $t_c(1)=0.5t$).  
Namely, there are two right-moving and 
two left-moving branches in the terminology of the weak-coupling 
Tomonaga-Luttinger (TL) theory.  
When the Umklapp processes are absent, 
the situation is, as far as the weak-coupling picture is concerned, 
essentially identical to the two-leg 
Hubbard ladder, where a spin gap opens when the two 
Fermi velocities are not too different. 
Thus a spin gap
also opens in the present model for a certain range of the ratio of the 
Fermi velocities, i.e., for $|t'|>t_c'(n)(>t_c(n))$. 
In the presence of a spin gap, the 
pairing correlation dominates for small enough $U$ 
in the two-leg ladder when doped (i.e., at non-half-filling),
\cite{Fabrizio,Schulz,Balents,Kuroki} 
and this is also the case with the present NNN model. 

An important difference, however, appears at half-filling, where Umklapp
processes emerge. In the two-leg ladder, in which the Umklapp process is a
two-electron scattering process at half-filling,   
an infinitesimal $U$ is enough to make the Umklapp process relevant in 
the renormalization, leading to an opening of a charge gap.  
In contrast, Umklapp processes are only higher-order 
at half-filling in the present NNN model, in which 
a four-electron scattering is involved.

The dominant pairing considered in Fabrizio's weak-coupling picture is 
\begin{equation}
c_{k_F^1\uparrow}c_{-k_F^1\downarrow}-c_{k_F^2\uparrow}c_{-k_F^2\downarrow},
\label{order}
\end{equation}
which has the correlation function of the form 
$\sim 1/r^{1/2K}$ ($r$: the real
space distance).  This pairing dominates 
over a charge-density wave (or dimer wave in Fabrizio's terminology) 
whose correlation is $\sim 1/r^{2K}$.  
If we assume that the analysis may be extended to finite $U$'s, 
the renormalization equation has 
such a structure that 
$K\geq 0.5$ for $U\leq U_C$ 
(with $K\rightarrow 1$ for $U\rightarrow +0$), where 
the pairing correlation is the most slowly-
decaying ($\sim 1/r^{0.5}$ in the weak-coupling limit as in 
doped two-\cite{Fabrizio,Schulz,Balents} and three-leg
\cite{Schulz,Kimura} Hubbard ladders).  
As $U$ is increased the Umklapp process becomes relevant 
to make the system insulating 
precisely when the two functional forms coincide 
($1/r^{1/2K} = 1/r^{2K}$, i.e., at $U=U_C$).

Now let us turn to our numerical analysis for finite $U$'s.  
We take $t'\simeq -0.8t$ as a typical value of $t'> t_c =0.5t$. 
We first present the
QMC result of the pairing correlation function in the 
ground state. We have adopted the projector Monte Carlo method, in which we
implement the stabilization procedure adopted by several 
authors.\cite{Sorella,White,Imada2} We have taken projecting time $\tau$ 
of up to $\sim 60/t$ with Trotter slices $L$ of $\tau/L\leq0.2/U$ 
to assure the convergence of the correlation function.
The details of the calculation is similar to that for 
the two-leg Hubbard ladder in ref.\onlinecite{Kuroki}
We take 
60 electrons on 60 sites with $t'=-0.78t$.  
For this set of parameters, 
the energy levels of the two branches of the one-electron band structure 
becomes very close to each other at the Fermi level. 
It has been revealed for two-\cite{Kuroki} and three-leg\cite{Kimura}
Hubbard ladders that
such an alignment is desirable to mimic infinite systems, and is 
indeed 
necessary in order to obtain correlation functions
consistent with the weak-coupling theory at large distances.  
This applies to the present model as well. 
As suggested in 
ref.\onlinecite{Kuroki}, this may be an indication that the offset between
the discrete energy levels has to be smaller than the spin gap in order to
obtain results consistent with the weak-coupling theory in finite size
systems.

For this parameter set the Fermi wave numbers satisfy a relation 
$k_F^1-k_F^2=\pi/2$, so that the relevant pairing correlation 
becomes 
NNN pairs in real space according to eqn.\ref{order}, since 
$|\exp(ik_F^1\Delta x)-\exp(ik_F^2\Delta x)|$ takes its maximum at 
$\Delta x=2$. 
Thus we study the pairing correlation, 
\begin{eqnarray}
&&P(r)\equiv \langle O_{i+r}^\dagger O_{i}\rangle\nonumber\\
&&O_i=(c_{i\uparrow} c_{i+2\downarrow}-c_{i\downarrow} 
c_{i+2\uparrow})/\sqrt{2}.
\end{eqnarray}

\begin{figure}
\begin{center}
\leavevmode\epsfysize=70mm \epsfbox{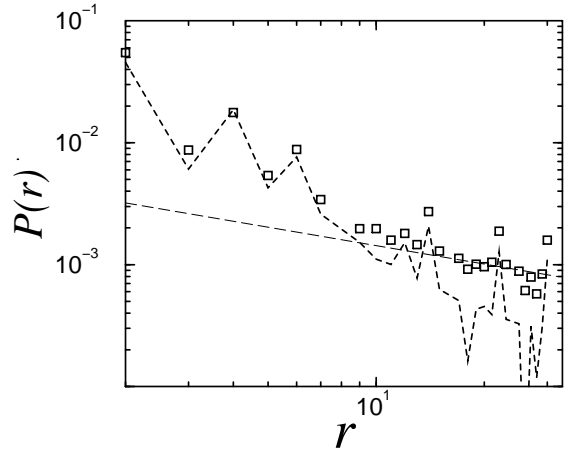}
\caption{
QMC result for
the pairing correlation function, $P(r)$, plotted 
against the real space distance $r$ in a 60-site 
half-filled Hubbard ladder with $U=t$ and $t'=-0.78t$($\Box$).
The dashed line is the noninteracting
result for the same system size, while the straight dashed line 
represents $\sim 1/r^{0.5}$. The pairing correlation function
takes negative values at distances $r=8,16,24$, which are omitted in the
plot.
}
\end{center}
\label{fig1}
\end{figure}

Figure \ref{fig1} shows the pairing correlation function for 
a periodic 60-site system with $U=t$. The pairing correlation is clearly
enhanced over the non-interacting case, and a slowly decaying component 
close to $\sim 1/r^{0.5}$ can be seen at large distances. 
The behavior is similar to that for two-\cite{Kuroki} and 
three-leg\cite{Kimura} Hubbard ladders.

\begin{figure}
\begin{center}
\leavevmode\epsfysize=70mm \epsfbox{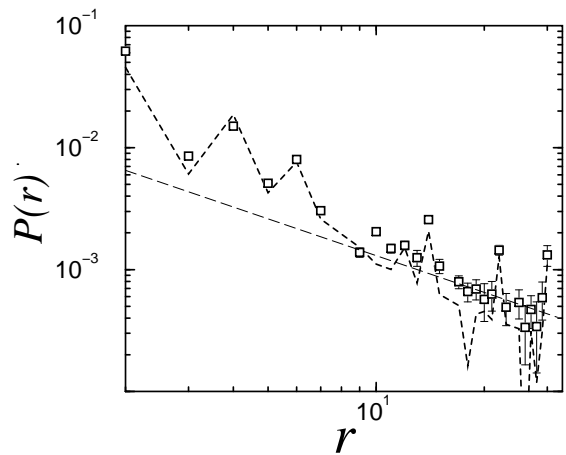}
\caption{
A plot similar to Fig.1 for $U=2t$ and $t'=-0.78t$.
The straight dashed line represents $\sim 1/r$. 
}
\end{center}
\label{fig2}
\end{figure}

Figure \ref{fig2} is a similar plot for a larger $U=2t$, 
about the largest $U$ for which the negative-sign problem 
in QMC is overcome.  
In this case, the enhancement is smaller than that for $U=t$, 
which suggests that the exponent $K$ is indeed a decreasing function of $U$.  
However, the existence of the enhancement itself implies 
that the charge gap is still closed there, because otherwise the 
pairing correlation would decay exponentially. 
The exponent at large distances is seen to be close to 
unity, which indicates a proximity to the superconductor-insulator 
transition, so that 
$U_C$ should lie 
somewhere above, but not too far from, $U=2t$.  

The regime with larger $U$ is exactly where we can introduce the 
DMRG method.\cite{White2} 
The calculation has 
been done for system sizes up to $L=28$ sites 
with $t'=-0.8t$ here with an open boundary condition. 
We have kept up to 120 states per block with truncation 
errors smaller than $10^{-4}$ and mostly around $10^{-5}\sim 10^{-6}$.
The charge gap $\Delta_C(N)$ and the spin gap $\Delta_S(N)$ for a half-filled
$N$-site system are calculated by 
\begin{eqnarray}
\Delta_C(N)&=&\left[E(N/2+1,N/2+1)
+E(N/2-1,N/2-1)\right.\nonumber\\
&&\left.-2E(N/2,N/2)\right]/2,\\
\Delta_S(N)&=&E(N/2+1,N/2-1)-E(N/2,N/2),
\end{eqnarray}
respectively, where $E(N_\uparrow,N_\downarrow)$ is the ground state
energy for $N_\uparrow$ up-spin and $N_\downarrow$ down-spin 
electrons.
We have checked that as far as these quantities and the 
parameter values adopted here are 
concerned, $\sim 100$ states per block suffice for the convergence.\cite{corr}

\begin{figure}
\begin{center}
\leavevmode\epsfysize=70mm \epsfbox{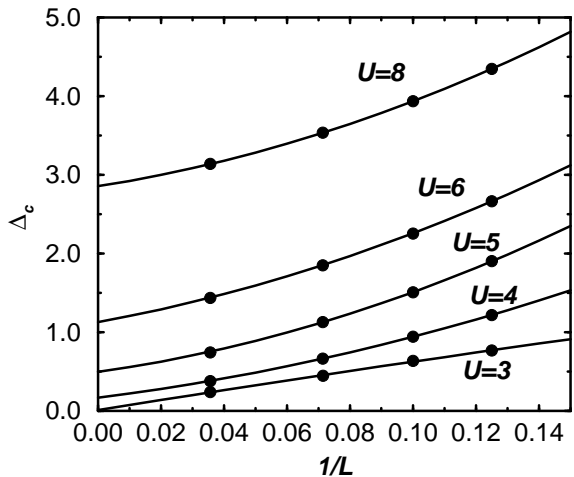}
\caption{
DMRG evaluation of the charge gap $\Delta_C$ at $t'=-0.8t$
plotted against $1/L$ for various values of $U$. 
Dashed curve is a least-squares fit with second
order polynomials in $1/L$.
}
\end{center}
\label{fig3}
\end{figure}

Figure \ref{fig3} displays the charge gap for various
values of $U$ as a function of $1/L$ (the inverse system size).
The results for 8- and 10-site systems are also obtained from the exact
diagonalization in order to check the DMRG result.  
The system-size dependence is least-squares fit to second-order 
polynomials in $1/L$. 
The extrapolation to $L\rightarrow\infty$ shows that the charge
gap closes at $U_C\sim 3t$. 
Although the estimated $U_C$ 
may contain some errors due to finite-size effects, 
this result is consistent with the above QMC result that the 
pairing correlation exists for $U=2t$, which serves as 
a lower boundary for the existence of a spin gap.  
Thus the complementary QMC and DMRG results can indeed be combined to 
indicate a superconductor-insulator 
transition somewhere around $2t < U < 3t$.

\begin{figure}
\begin{center}
\leavevmode\epsfysize=70mm \epsfbox{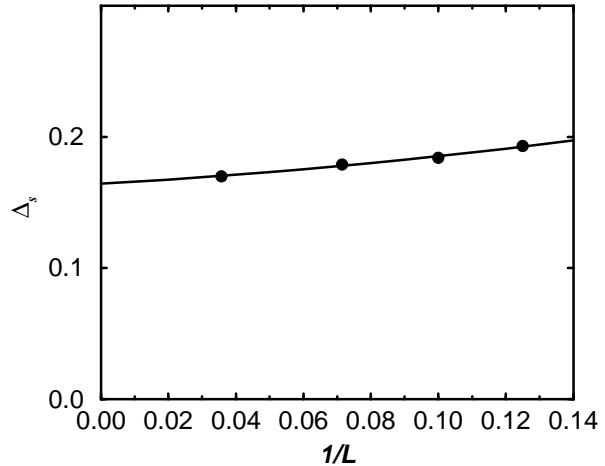}
\caption{
DMRG evaluation of the spin gap for $U=8t$ and $t'=-0.8t$ 
plotted against $1/L$.
}
\end{center}
\label{fig4}
\end{figure}
The existence of a metallic phase at half-filling having a spin gap and a 
dominating pairing correlation is rather surprising.  
One might be tempted to explain the origin of the spin gap and 
superconductivity by regarding the model as essentially the same as
the `$t$-$J$-$J'$' model,\cite{Ogata} 
where electrons hop with excluded double occupancies and 
interact with NN and NNN antiferromagnetic exchanges.  
This view, however, cannot be valid, 
because we are sitting on the half-filled point, 
where the $t$-$J$-$J'$ model is trivially insulating.  
In this sense the spin gap in the metallic phase of the present model 
may not be regarded as simply arising from a 
frustration in antiferromagnetic exchange interactions 
($J,J'$), which is consistent with 
Fabrizio's remark in ref.\onlinecite{Fabrizio2} on 
the inverse doping effects on the spin gap between the present model and the 
$t$-$J$-$J'$ model.

Finally we evaluate the spin gap using DMRG.  
Although the spin gap is a key concept in our motivation mainly 
in the metallic phase, 
we present here the result for the
insulating phase, which is of greater interest in the context of 
strongly correlated systems such as 1D cuprates.
The study of the spin gap in the metallic phase will be presented 
elsewhere.
We have evaluated the magnitude of the spin gap at $U=8t$ with the 
same value of $t'$ as above.
In fig.\ref{fig4}, we plot $\Delta_S$ as a function of $1/L$.
The extrapolation to the thermodynamic limit suggests an existence of a
spin gap as large as $\sim 0.16t\sim 0.17t$.\cite{White3}

To summarize, 
a transition from a metallic state with a dominant pairing 
correlation into an insulating state 
has been shown to occur at an intermediate strength of 
the interaction of $\sim O(t)$. 
Further studies for other values of $t'$ and $U$, and for the 
doped phase is under way.

We wish to thank Dr Y. Nishiyama for useful comments concerning
the DMRG method. We would also like to thank Dr K. Kusakabe and Dr T. Kimura 
for valuable discussions. 
Numerical calculations were done on FACOM VPP 500/40 at the Supercomputer 
Center, Institute for Solid State Physics, University of Tokyo, and 
HITAC S3800/280 at the Computer Center of the University of Tokyo.
This work was also supported in part by Grant-in-Aid for Scientific
Research from the Ministry of Education 
of Japan. 


\begin{figure}
\end{figure}


\end{multicols}

\begin{references}
\bibitem{corr} If we calculate correlation functions or the spin gap
in the metallic phase using DMRG, 
more states have to be kept as in ref.\onlinecite{Noack}, where 
400 states or more per block are kept.
\bibitem{White3} This may be compared with 
the $J$-$J'$ {\it Heisenberg} model
(R.Chitra et al, Phys. Rev. B {\bf 52}, 6581 (1995); 
S.R. White and I. Affleck, Phys. Rev. B{\bf 54}, 9862 (1996)), 
for which the magnitude of the spin gap is 
evaluated with DMRG to be $\sim 0.42J$
for $J'\sim(0.8)^2J$.  If we assume $J=4t^2/U$ and
$J'=4t'^2/U$, $0.42J$ amounts to $0.21t$ for $U=8t$.
\end{references}
\end{document}